\documentclass[a4paper,11pt]{article}%
\usepackage{amssymb,amsmath,amsfonts}
\usepackage{graphicx}
\usepackage{amsmath}
\usepackage{amsfonts}
\usepackage{amssymb}%
\providecommand{\U}[1]{\protect\rule{.1in}{.1in}}

\newtheorem{idea memo}[theorem]{Idea Memo}

\begin{document}

\title{Space(-Time) Emergence as \\Symmetry Breaking Effect\thanks{Talk at ICQBIC (10 -- 13 March 2010) at Tokyo
University of Sciences, Noda }}
\author{Izumi OJIMA}
\date{}
\maketitle

\begin{abstract}
The microscopic origin of space(-time) geometry is explained on the basis of
an emergence process associated with the condensation of infinite number of
microscopic quanta responsible for symmetry breakdown, which implements the
basic essence of \textquotedblleft Quantum-Classical
Correpondense\textquotedblright\ and of the forcing method in physical and
mathematical contexts, respectively. From this viewpoint, the space(-time)
dependence of physical quantities arises from the \textquotedblleft logical
extension\textquotedblright\ \cite{OjiOza} to change \textquotedblleft
constant objects\textquotedblright\ into \textquotedblleft variable
objects\textquotedblright\ by tagging the order parameters associated with the
condensation onto \textquotedblleft constant objects\textquotedblright; the
logical direction here from a value $y$ to a domain variable $x$ (to
materialize the basic mechanism behind the Gel'fand isomorphism) is just
opposite to that common in the usual definition of a function $f:x\longmapsto
f(x)$ from its domain variable $x$ to a value $y=f(x)$.

\end{abstract}

\section{Outline of the Problem}

Before going into the main context, a comment would be necessary on the
parenthesis in \textquotedblleft Space(-Time) Emergence\textquotedblright\ in
the title: in sharp contrast to the case of \textit{space}, the emergence of
\textit{time axis} seems now doubtful. To justify this suspicion, we need
re-examine its consistency with the \textit{space-time picture }essential in
special and general theories of relativity, which is not undertaken yet here.
This is the reason for the expression \textquotedblleft
Space(-Time)\textquotedblright.

\subsection{\textquotedblleft Theory of Everything\textquotedblright\ vs.
Duheme-Quine thesis}

In search for a new theory to incorporate the old and standard one as a
special case, one usually attempts trial-and-error searches in a
\textit{heuristic} way which seems to be unavoidable. How and to which extent
can this be made systematic by the method for solving \textquotedblleft%
\textit{\textbf{\textit{inverse problem}}}\textquotedblright?~In this context,
we note the existence of an obstruction to this possibility in such a form as
\textquotedblleft Duheme-Quine thesis\textquotedblright. This is just a
\textit{\textbf{\textit{No-Go theorem}}} telling the impossibility to
determine uniquely a theory from phenomenological data so as to reproduce the
latter, because of unavoidable \textit{\textbf{\textit{finiteness}}}
in\textit{\ \textbf{\textit{number}}} of measurable quantities and of their
\textit{\textbf{limited accuracy}}: 
\[%
\begin{array}
[c]{ccc}%
\begin{array}
[c]{c}%
\text{\textit{\textbf{non-unique}}~choices}\\
\text{of starting \textquotedblleft\textit{\textbf{Micro}}\textquotedblright:}%
\end{array}
&
\begin{array}
[c]{c}%
\text{predictions: \textit{\textbf{\textit{not 1-to-1}}}}\\
\rightleftarrows\\
\text{inference: \textit{\textbf{\textit{not onto}}}}%
\end{array}
&
\begin{array}
[c]{c}%
\text{\textit{\textbf{finite data }}with}\\
\text{\textit{\textbf{limited}} \textit{\textbf{accuracy}}}\\
\text{at \textquotedblleft\textit{\textbf{Macro}}\textquotedblright\ level}%
\end{array}
\end{array}
.
\]
\newline Owing to \textit{\textbf{\textit{inevitable }errors}} in the measured
data, the agreements between theoretical predictions and experimental data
justify the former only as \textit{\textbf{one of the possible candidates}} to
explain the latter: 

\[%
\begin{array}
[c]{ccc}%
\text{Theory 1} & {\searrow} & \\
\text{Theory 2} & \longrightarrow & \text{Experimental data}\mathcal{~}%
{+~}\text{{errors}}\\
\mathcal{\vdots} & {\nearrow} &
\end{array}
.
\]

Fortunately, our \textit{\textbf{bi-directional }}method of \textquotedblleft%
\textit{\textbf{\textit{Micro-Macro duality}}}\textquotedblright%
\ \cite{MicroMacro} \textit{\textbf{\textit{can}}}
\textit{\textbf{\textit{resolve}}} this \textit{\textbf{\textit{universal
dilemma }}}in harmony with the necessary and sufficient levels of accuracy
determined by the inevitable restrictions on \textit{\textbf{\textit{focused
aspects}} }and\textit{\ \textbf{\textit{degrees of accuracy}}} inherent in a
certain pre-chosen \textit{\textbf{\textit{context}}}. Within such a context,
a theoretical explanation can be unicified by \textquotedblleft%
\textit{\textbf{\textit{Micro-Macro duality}}}\textquotedblright\ as a
context-dependent \textquotedblleft\textit{\textbf{matching condition}%
}\textquotedblright\textit{\textbf{\ }}\cite{Unif03} between the phenomena
(\textquotedblleft Macro\textquotedblright) to be described and the theory to
describe (\textquotedblleft Micro\textquotedblright); \textquotedblleft%
\textit{\textbf{Macro}}\textquotedblright\ in this mathematical formulation
plays the roles of a \textit{\textbf{\textit{standard reference frame}}}
characterized by its \textquotedblleft\textit{\textbf{\textit{universality}}%
}\textquotedblright\ (in the mathematical and categorical sense). This
naturally leads us to the idea of \textquotedblleft\textit{\textbf{matching
condition}}\textquotedblright\ between \textit{\textbf{inductive }}\&
\textit{\textbf{deductive}} aspects\textit{\textbf{\ }}for judging the
correctness, which demarcates and characterizes the target domain of discourse.

\subsection{\textquotedblleft Geometrization principle\textquotedblright\ vs.
Physical emergence of space-time}

In contrast to the above resolution of Duheme-Quine thesis by Micro-Macro
duality:
\[
Micro\overset{\text{deduction}}{\underset{\text{induction}}{\rightleftarrows}%
}\ Macro,
\]
the \textquotedblleft standard\textquotedblright\ approach regards the
\textquotedblleft rigorous\textquotedblright\ deductions from Micro to Macro
as the only possible scientific paths to be followed. In this context, the
starting point of Micro Theory consists simply of \textit{\textbf{\textit{ad
hoc postulates}}} which cannot be justified within a theory itself, to be
justified experimentally up to certain limited accuracy. With this point
neglected, however, theoretical hypotheses on Micro quantum systems are always
\textit{absolutized}, in combination with the basic principle of
\textquotedblleft\textit{\textbf{\textit{geometrization of physics}}%
}\textquotedblright\ prevailing in modern physics. However, we need to ask
what its foundation is: it turns out to be based upon the successes of methods
of modern geometry in mathematics and in its physical applications, such as
general relativity, gauge theories, etc., which are essentially
\textit{\textbf{of macroscopic nature}}!! Almost all the basic principles
governing modern geometries (differential, complex-analytic, algebraic, etc.)
have been extracted from and applied to \textit{\textbf{classical \&
macroscopic levels }}based on commutative algebras (of observables), and its
\textit{\textbf{quantum versions}} have only started to be sought for, without
reaching mature stages yet!

In spite of strong emphasis there on the \textit{rigorous} derivations of
Macro(-scopically observable predictions) from Micro, \textit{\textbf{neither
}}the origin of \textit{\textbf{space-time as Macro, nor }}\textquotedblleft%
\textit{\textbf{\textit{\textbf{\textit{Macro }}principle of geometrization}}%
}\textquotedblright\ seem to be well founded!? These pitfalls at both the
Micro and Macro ends seem to be the two fatal defects of the fashionable
trends in modern physics hidden in its blind spots. Therefore, we need explain
the \textit{\textbf{microscopic origins of macroscopic structures of
\textit{space-time geometry itself}}}. This is just the problem to search the
physical origin and emergent processes of spacetime structure in microscopic
physics, to be pursued in the following.

\section{Universality inherent in Macro-levels}

For this latter purpose also, the methods based on \textquotedblleft
Micro-Macro duality\textquotedblright\ will turn out to be quite effective, as
shown below. In this context, what plays the most crucial roles can be found
in the construction of a \textit{\textbf{Micro-Macro composite system}}
consisting of Micro and Macro levels based upon the
\textit{\textbf{\textit{duality}}} between the two directions, deduction and induction:

i) deduction [Micro $\Longrightarrow$ Macro]= a
\textit{\textbf{\textit{bundle}}} structure 
\[
\mathcal{A}\overset{i}{\mathcal{\hookrightarrow}}\mathcal{F}\overset
{p}{\twoheadrightarrow}\mathcal{F}/\mathcal{A}\simeq\widehat{Gal(\mathcal{F}%
/\mathcal{A})}%
\]
(formulated as an exact sequence, $\operatorname{Im}i=\ker
p$, of \textquotedblleft triples\textquotedblright\ equipped with a tri-linear
multiplication, e.g., $\mathcal{A}\times\mathcal{A}\times\mathcal{A}%
\ni(A,B,C)\longmapsto\{A,B,C\}\in\mathcal{A}$, depending linearly on $A$ and
$C$ and anti-linearly on $B$) and

ii) [Micro $\Longleftarrow$ Macro] induction = the corresponding
\textit{\textbf{connections}} defined as its \textit{\textbf{splittings}%
},\textit{\textbf{\ }}$\mathcal{A}\overset{m}{\mathcal{\twoheadleftarrow}%
}\mathcal{F}\overset{h}{\hookleftarrow}\mathcal{F}/\mathcal{A}$, characterized
by one of the mutually equivalent three conditions, 
\[
m\circ i=1_{\mathcal{A}},p\circ h=1_{\mathcal{F}/\mathcal{A}},i\circ m+h\circ
p=1_{\mathcal{F}}.
\]
\vskip12pt Example: The first law of thermodynamics describes a dilation
$\Delta E=Q+W$ of \textit{\textbf{heat}} $Q$ and of \textit{\textbf{work}} $W$
into a closed dynamical system with conserved energy\textit{\textbf{ }}$\Delta
E$. Here, the heat $Q$ symbolically represents the
\textit{\textbf{\textit{uncontrollable}}} component in Macro-manifestation of
\textit{\textbf{invisible}} Micro-motions as \textit{\textbf{holonomy }}in the
thermal classifying space (consisting of the basic thermodynamic order
parameters) and the work $W$ Macro-aspects of thermal system directly
\textit{\textbf{controllable at Macro-level}}, both of which are unified via
dilation into \textit{\textbf{Micro\textit{-Macro composite system}}} as a
\textit{\textbf{closed dynamical system}} with \textit{\textbf{conserved
energy}}. This relation can be expressed concisely in terms of the exact
sequence in such a form as [Micro \textit{\textbf{fibre}}:\textit{\textbf{\ }%
}$Q$]$\overset{i}{\hookrightarrow}$[$\Delta E=Q+W$:
\textit{\textbf{Micro\textit{-Macro composite system}}}%
]\textit{\textbf{\textit{ }}}$\overset{p}{\twoheadrightarrow}$[$W$: Macro
\textquotedblleft\textit{\textbf{base space}}\textquotedblright], where
$\operatorname{Im}i\subset\ker p$ means that heat$\in\operatorname{Im}i$
cannot be transformed into controllable work as its projection by $p$ equals
$0$, and, conversely, $\ker p\subset\operatorname{Im}i$ means that any energy
$\in\ker p$ unchangeable to work should be regarded as heat $\in
\operatorname{Im}i$. Thus, the \textit{\textbf{bundle structure}} +
\textit{\textbf{exactness }}can be seen to carry relevant physical or
operational meanings. \vskip10pt Other interesting examples can also be found,
for instance, in the theories of Maxwell and of Einstein in such forms as:
\vskip10pt
\[%
\begin{array}
[c]{ccc}%
\text{Micro} & \text{Micro-Macro} & \text{Macro}\\
J_{\mu} &  & F_{\mu\nu}\\
\uparrow\downarrow & \overset{\text{Maxwell Eqn}}{\underset
{\text{Electromagmetoc Forces}}{\rightleftarrows}} & \uparrow\downarrow\\
&  & \\
\psi &  & A_{\mu}%
\end{array}
\]
\vskip10pt

\noindent and
\[%
\begin{array}
[c]{ccc}%
\text{Micro} & \text{Micro-Macro} & \text{Macro}\\
T_{\mu\nu} &  & R_{\mu\nu}\\
\uparrow\downarrow & \overset{\text{Einstein Eqn}}{\underset
{\text{Gravitational Force}}{\rightleftarrows}} & \uparrow\downarrow\\
&  & \\
\psi &  & \Gamma_{\mu\nu}^{\lambda}%
\end{array}
.
\]
\vskip12pt In many cases the above exact sequence takes such a form that
$\mathcal{A}$ and $\mathcal{F}$ are (C*-)algebras, and the triple
$\mathcal{F}/\mathcal{A}$ can be viewed as $\mathcal{A}$-module $\mathcal{F}$
controlled by Galois group $G=Gal(\mathcal{F}/\mathcal{A})$ defined by such a
subgroup of automorphisms $Aut(\mathcal{F})$ of $\mathcal{F}$ as consisting of
elements $\in G$ fixing $\mathcal{A}=\mathcal{F}^{G}$ pointwise.

In this case,$\ \widehat{Gal(\mathcal{F}/\mathcal{A})}=\hat{G}$ can be
regarded as the totality of irreducible unitary representations of $G$ (if
such is meaningful) or the tensor category consisting of unitary
representations of $G$ and the map $p:\mathcal{F}\twoheadrightarrow\widehat
{G}$ extracts the $G$-representation contents of each element in $\mathcal{F}%
$. If we equip $\mathcal{F}$ with an $\mathcal{A}$-valued inner product,
$\mathcal{F}\times\mathcal{F}\ni(F_{1},F_{2})\longmapsto\langle F_{1}%
|F_{2}\rangle\in\mathcal{A}$, it becomes a Hilbert module with a right action
of $\mathcal{A}$, and a splitting,\textit{\textbf{\ }}$\mathcal{A}\overset
{m}{\mathcal{\twoheadleftarrow}}\mathcal{F}\overset{h}{\hookleftarrow
}\mathcal{F}/\mathcal{A}=\hat{G}$, can be specified by the conditional
expectation value $\mathcal{A}\overset{m}{\mathcal{\twoheadleftarrow}%
}\mathcal{F}$ arising from an $\mathcal{A}$-valued inner product of
$\mathcal{F}$ by $m(F)=\langle1|F\rangle$ if $1\in\mathcal{F}$ (or considering
an approximate unit of $\mathcal{F}$). Then, $\mathcal{F}$ can be recovered
from $\mathcal{A}$ and $\hat{G}$ as a Galois extension by a crossed product:
$\mathcal{F}=\mathcal{A}\rtimes\hat{G}$ which gives a typical example of
\textit{\textbf{dilation }}from Macro to Micro.

In this way, the duality between bundle structure $\mathcal{A}\overset
{i}{\mathcal{\hookrightarrow}}\mathcal{F}\overset{p}{\twoheadrightarrow
}\mathcal{F}/\mathcal{A}\simeq\widehat{Gal(\mathcal{F}/\mathcal{A})}$ and its
connection $\mathcal{A}\overset{m}{\mathcal{\twoheadleftarrow}}\mathcal{F}%
\overset{h}{\hookleftarrow}\mathcal{F}/\mathcal{A}$ can be seen to condense
the essence of \textit{\textbf{Fourier-Galois duality}}, especially because
the functors $Gal$ and $G\longmapsto\hat{G}$ assign, respectively, a group
$Gal(\mathcal{F}/\mathcal{A})$ to the $\mathcal{A}$-module $\mathcal{F}%
/\mathcal{A}$ and the representation contents $\hat{G}$ of $G$ to a group $G$.

Extending this Fourier-Galois theoretical machinery to a symmetry breaking
situation of a $G$-dynamical system $\mathcal{F}\underset{\tau}%
{\curvearrowleft}G$ with a fixed-point subalgebra $\mathcal{A}=\mathcal{F}%
^{G}$, we see below that a process of space(-time) emergence can be formulated
as a kind of \textit{\textbf{symmetry breaking}} in terms of the notions of an
\textit{\textbf{augmented algebra}} and of an associated
\textit{\textbf{sector bundle }}\cite{Unif03}.

\section{\textquotedblleft Sector bundle\textquotedblright\ associated with
broken symmetry}

Breakdown of a symmetry of $\mathcal{F}$ in a state $\omega\in E_{\mathcal{F}%
}$ with a group $G$ into a subgroup $H\subset G$ of a remaining symmetry is
characterized \cite{Unif03} by the non-invariance of the \textquotedblleft
central extension\textquotedblright\ of $\omega$ on the centre $\mathfrak{Z}%
_{\pi_{\omega}}(\mathcal{F}):=\pi_{\omega}(\mathcal{F})^{\prime\prime}\cap
\pi_{\omega}(\mathcal{F})^{\prime}$ under the corresponding $G$-action on
$\mathfrak{Z}_{\pi_{\omega}}(\mathcal{F})$. In this situation, the role of
algebra $\mathcal{A}=\mathcal{F}^{G}$ of observables is known in algebraic QFT
to be replaced by the Haag-dual extension $\mathcal{A}^{d}$ owing to the
breakdown $\mathcal{A\subsetneqq A}^{d}$ of Haag duality, where the Haag-dual
net $\mathcal{A}^{d}$ is defined with respect to a vacuum representation
$\pi_{0}$ by $\mathcal{A}^{d}(\mathcal{O}):=[\pi_{0}^{-1}](\pi_{0}%
(\mathcal{A}(\mathcal{O}^{\prime}))^{\prime})$ (for $\forall\mathcal{O}$:
double cones in Minkowski spacetime), so that the sector structure is
determined by the factor spectrum $\overset{\frown}{\mathcal{A}^{d}%
}=Spec(\mathfrak{Z}(\mathcal{A}^{d}))=\hat{H}$: the group dual of a compact
Lie group $H$ consisting of its irreducible unitary representations:
$\mathcal{A}$ $\Longrightarrow\mathcal{A}^{d}=\mathcal{F}^{H}$ ($\mathcal{F}%
=\mathcal{A}^{d}\rtimes\widehat{H}$).

A general and desirable definition of \textit{\textbf{the}} group $G$ of
\textit{\textbf{broken }}symmetry is not known yet in terms of the above data
coming from the Haag dual net $\mathcal{A}^{d}$, but such a definition as
$G:=Gal(\mathcal{F}/\mathcal{A})$ with the field algebra $\mathcal{F}%
=\mathcal{A}^{d}\rtimes\widehat{H}\rightleftarrows\mathcal{A}^{d}%
=\mathcal{F}^{H}$ and the group of unbroken symmetry: $H=Gal(\mathcal{F}%
/\mathcal{A}^{d})\subset G$, is sufficient for our purposes when the obtained
$G$ is a finite-dimensional Lie group.

With $\mathcal{\tilde{F}}:=\mathcal{A}^{d}\rtimes\widehat{G}=\mathcal{F}%
\rtimes\widehat{(H\backslash G)}$ called an \textit{\textbf{augmented algebra
}}\cite{Unif03}, we have a \textit{\textbf{split }}bundle exact sequence
$\mathcal{A}^{d}\overset{\tilde{m}}{\underset{\mathcal{\hookrightarrow}%
}{\twoheadleftarrow}}\mathcal{\tilde{F}}\underset{\twoheadrightarrow
}{\hookleftarrow}\mathcal{\tilde{F}}/\mathcal{A}^{d}\simeq\widehat{G}$. In
this situation, the \textit{\textbf{minimality}} of\textit{\textbf{\ }}$G$ and
$\mathcal{\tilde{F}}$\ is guaranteed by the $G$-\textit{\textbf{central
ergodicity}}, i.e., $G$-ergodicity of the centre $\mathfrak{Z}_{\tilde{\pi}%
}(\mathcal{\tilde{F}})$ in the representation $\tilde{\pi}$ given by the GNS
representation of $\omega_{0}\circ\tilde{m}$ induced from the vacuum state
$\omega_{0}$ of $\mathcal{A}^{d}$ \cite{Unif03}, and we have the following
commutativity diagram: 
\[%
\begin{array}
[c]{ccc}
& \mathcal{F}^{H}=\widetilde{\mathcal{F}}^{G}\text{: }%
\begin{array}
[c]{c}%
\text{unbroken alg.}\\
\text{of observables}%
\end{array}
& \\
^{\text{1:1}}\swarrow &  & \searrow^{\text{1:1}}\text{ \ \ \ \ \ }\\
\mathcal{F} &  & \widetilde{\mathcal{F}}^{H}\text{: extended observables}\\
& \searrow\searrow^{\text{1:1}}\text{\ \ \ \ \ \ }\Downarrow^{\text{1:1}%
}\text{ \ \ \ \ \ \ \ \ \ \ \ }^{\text{1:1}}\swarrow & \\
^{\text{onto}}\downarrow & \text{ \ \ \ \ \ \ \ \ \ \ \ \ \ \ }\widetilde
{\mathcal{F}}\text{: augmented alg.\ \ } & \text{\ }\downarrow^{\text{onto}}\\
^{\text{onto}}\downarrow & \swarrow^{\text{onto}}\text{\ \ \ \ \ }%
\Downarrow^{\text{onto}}\text{ \ \ \ \ }^{\text{onto}}\searrow\searrow &
\text{\ }\downarrow^{\text{onto}}\\
\text{ \ \ \ \ }\widehat{H} & \twoheadleftarrow\text{ \ \ \ \ \ }\widehat
{G}\text{ \ \ \ \ \ \ \ \ \ \ \ \ \ }\hookleftarrow & \text{\ }\widehat
{G/H}\text{ \ \ \ \ \ }%
\end{array}
,
\]
whose dual version describes the sector structure: 

\[%
\begin{array}
[c]{ccc}
& \overset{\frown}{\widetilde{\mathcal{F}}^{G}}=\text{ }\overset{\frown
}{\mathcal{F}^{H}}\simeq\text{\ }\widehat{H} & \text{: unbroken\ sectors}\\
& \nearrow^{\text{onto}}\text{\ \ \ }\Uparrow\text{\ \ \ \ \ \ \ \ }%
\nwarrow^{\text{onto}}\text{ \ \ \ \ \ } & \Downarrow^{\text{1:1}}\\
\overset{\frown}{\mathcal{F}} & \text{\ \ \ \ \ \ \ \ \ \ \ \ }\Uparrow
^{\text{onto}}\text{\ \ \ }\overset{\frown}{\widetilde{\mathcal{F}}^{H}}\simeq
G\underset{H}{\times}\widehat{H} & \text{: sector bdle \ \ \ \ \ \ }\\
^{\text{1:1}}\uparrow & \nwarrow\nwarrow^{\text{onto}}\Uparrow
\text{\ \ \ \ \ \ \ }\nearrow^{\text{onto}}\text{\ \ \ \ }\uparrow &
\Downarrow\text{ \ \ \ }\\
^{\text{1:1}}\uparrow & \text{\ \ \ \ \ \ \ \ \ \ \ }\overset{\frown
}{\widetilde{\mathcal{F}}}\text{ \ \ \ \ \ \ \ \ \ \ \ \ \ \ \ \ \ }%
^{\text{1:1}}\uparrow & \Downarrow^{\text{onto}}\\
^{\text{1:1}}\uparrow & \nearrow^{\text{1:1}}\text{\ \ \ \ }\Uparrow
^{\text{1:1}}\text{\ \ \ \ }\nwarrow\nwarrow^{\text{1:1}}\text{\ \ }\uparrow &
\Downarrow\text{ \ \ \ }\\
H & \hookrightarrow\text{\ \ \ \ \ \ }G\text{: broken\ \ \ \ }%
\twoheadrightarrow\text{ \ \ }G/H & \text{: degenerate vacua\ \ \ \ \ \ \ }%
\end{array}
,
\]
where $\overset{\frown}{\mathcal{F}}=Spec(\mathfrak{Z}%
(\mathcal{F}))$ denotes the factor spectrum of $\mathcal{F}$, etc. \vskip10pt
\textit{\textbf{Remark}}: The physical essence of the extension $\mathcal{A}%
\Longrightarrow\mathcal{A}^{d}$ from the original observable algebra
$\mathcal{A}$ to its Haag-dual net algebra $\mathcal{A}^{d}=\mathcal{F}^{H}$
can now be interpreted as an \textquotedblleft extension of coefficient
algebra\ $\mathcal{A}$\textquotedblright\ by (the dual of) $G/H$ to
parametrize the degenerate vacua: $\mathcal{A}^{d}=\mathcal{F}^{H}%
=\widetilde{\mathcal{F}}^{G}=[(\mathcal{F}\rtimes\widehat{(H\backslash
G)}]^{G}=\mathcal{F}^{G}\rtimes\widehat{(H\backslash G)}=\mathcal{A}%
\rtimes\widehat{(H\backslash G)}$. In this extension, a part $G/H$~of
originally \textit{\textbf{invisible }}$G$ becomes \textit{\textbf{visible }%
}through the \textit{\textbf{emergence of degenerate vacua}} parametrized
by\textit{\textbf{\ }}$G/H$\textit{\textbf{\ }}due to the
\textit{\textbf{condensation of order parameters }}$\in G/H$ associated with
\textbf{S}(ponteneous) \textbf{S}(ymmetry) \textbf{B}(reaking) of $G$ to $H$.
As a result, observables $A\in\mathcal{A}$ acquire $G/H$-dependence:
$\widetilde{A}=(G/H\ni\dot{g}\longmapsto\widetilde{A}(\dot{g})\in
\mathcal{A})\in\mathcal{A}\rtimes\widehat{(H\backslash G)}$, which should just
be interpreted as an example case of \textit{\textbf{logical extension }%
}\cite{OjiOza} transforming a \textquotedblleft\textit{\textbf{constant}}
object\textquotedblright\ ($A\in\mathcal{A}$) into a \textquotedblleft%
\textit{\textbf{variable}} object\textquotedblright\ ($\widetilde{A}%
\in\mathcal{A}\rtimes\widehat{(H\backslash G)}$) having
\textbf{\textit{functional dependence}} on the
universal\textbf{\textit{\ classifying space}} $G/H$\ for (multi-valued)
\textbf{\textit{semantics}}(, as is familiar in the non-standard and
Boolean-valued analyses). By replacing $G/H$ with the space(-time), the above
consideration can be utilized as a prototype for the origin of the functional
dependence of physical quantities on space(-time) coordinates, due to the
physical emergence of space(-time) from microscopic physical world.

Along this line, we prescribe the similar logical extension procedure on the
observable algebra $\mathcal{A}^{d}=\mathcal{F}^{H}$ adding $G/H$-dependence:
\[
\mathcal{A}^{d}\rtimes\widehat{(H\backslash G)}=\mathcal{F}^{H}\rtimes
\widehat{(H\backslash G)}=(\mathcal{F}\rtimes\widehat{(H\backslash G)}%
)^{H}=\widetilde{\mathcal{F}}^{H}.
\]
Then, the whole sector structure of $\widetilde{\mathcal{F}}^{H}%
=(\mathcal{F}^{H}\rtimes\widehat{(H\backslash G)})$ can be identified with its
factor spectrum $\overset{\frown}{\widetilde{\mathcal{F}}^{H}}=G\underset
{H}{\times}\hat{H}$; this is seen to constitute a bundle structure, $\hat
{H}\hookrightarrow\overset{\frown}{\widetilde{\mathcal{F}}^{H}}=G\underset
{H}{\times}\hat{H}\twoheadrightarrow G/H$, called a \textit{\textbf{sector
bundle}} consisting of the classifying space $G/H$ of
\textit{\textbf{degenerate vacua}}, each fibre over which describes the sector
structure $\hat{H}$ corresponding to the \textit{\textbf{unbroken}} remaining
symmetry $H$ (or, more precisely, the conjugated group $gHg^{-1}$ for the
vacuum parametrized by $\dot{g}=gH\in G/H$).

Namely, the \textit{\textbf{sector bundle}}, $\hat{H}\hookrightarrow
\overset{\frown}{\widetilde{\mathcal{F}}^{H}}=G\underset{H}{\times}\hat
{H}\twoheadrightarrow G/H$, can be understood as the
\textit{\textbf{connection}}= \textit{\textbf{splitting}} of the dual,
$\overset{\frown}{\mathcal{F}^{H}}=\hat{H}\twoheadleftarrow\overset{\frown
}{\widetilde{\mathcal{F}}^{H}}=G\underset{H}{\times}\hat{H}\hookleftarrow
G/H$, of the bundle exact sequence of observable triples, $\mathcal{F}%
^{H}\hookrightarrow\widetilde{\mathcal{F}}^{H}=\mathcal{F}^{H}\rtimes
\widehat{(H\backslash G)}\twoheadrightarrow\widehat{(H\backslash G)}$!

\section{Emergence of space(-time) as symmetry breaking}

We can now apply the above scenario to the situation with the group $G$
describing both the external (= space-time) and the internal symmetries. For
simplicity, the latter component described by a subgroup $H$ of $G$ is assumed
to be unbroken, and hence, the broken symmetry described by $G/H$ represents
the space-time symmetry. It would be convenient to take $H$ as a normal
subgroup of $G$, though not essential. To be precise, $G/H$ may contain such
non-commutative components as spatial rotations (and Lorentz boosts) acting on
space(-time), we simply neglect this aspect to identify $G/H$ as the
space(-time) itself. Then, by identifying $G/H$ with a space(-time) domain
$\mathcal{R}$, we can notice a remarkable parallelism between the commutative
diagram in the previous section: 
\[%
\begin{array}
[c]{c}%
\mathcal{F}^{H}=\widetilde{\mathcal{F}}^{G}\\
_{H}\swarrow\mathcal{\ \ \ \ \ \ \ \ \ \ }\searrow_{G/H}\\
\mathcal{F}\text{\ \ \ \ \ \ \ \ \ \ }\Downarrow\text{\ \ \ \ \ \ \ \ }%
\widetilde{\mathcal{F}}^{H}\\
\text{$\downarrow$\ }_{G/H}\text{ }\searrow\searrow\text{\ \ \ }%
\mathcal{\ \ }\swarrow_{H}\text{ }\downarrow\\
\text{$\downarrow$\ }_{G/H}\text{ \ \ \ \ \ \ \ }\widetilde{\mathcal{F}%
}\text{\ \ \ \ \ \ \ \ \ \ \ }\downarrow\\
\downarrow\text{\ \ \ }\swarrow\text{\ \ \ }\Downarrow\text{ }\searrow
\searrow\text{\ \ \ }\downarrow\\
\widehat{H}\text{\ \ }\twoheadleftarrow\text{\ \ \ \ }\widehat{G}%
\text{\ \ }\hookleftarrow\text{ \ \ \ }\widehat{G/H}%
\end{array}
,
\]
and the diagram controlling Doplicher-Roberts
reconstruction of the local net $\mathcal{R}\longmapsto\mathcal{F}%
(\mathcal{R})$ from $\mathcal{R}\longmapsto\mathcal{A}(\mathcal{R})$:
\[%
\begin{array}
[c]{c}%
\mathcal{O}_{\rho}=O_{d}^{G}\\
_{G}\swarrow\mathcal{\ \ \ \ \ \ \ \ \ \ \ }\searrow_{\mathcal{R}}\\
\mathcal{O}_{d}\text{\ \ \ \ \ \ \ \ \ \ }\Downarrow\text{\ \ \ \ \ \ \ \ }%
\mathcal{A}(\mathcal{R})\\
\downarrow\text{ }_{\mathcal{R}}\searrow\searrow\text{ \ \ \ \ \ \ \ \ }%
\swarrow_{G}\text{ \ }\downarrow\\
\downarrow\text{ }_{\mathcal{R}}\text{ \ \ \ \ \ \ \ }\mathcal{F}%
(\mathcal{R})\text{\ }\ \ \ \ \ \ \ \ \text{ \ }\downarrow\\
\downarrow\text{ }\swarrow\text{\ \ \ \ \ \ }\Downarrow\text{\ \ \ \ \ }%
\searrow\searrow\text{\ }\downarrow\\
\widehat{G}\text{\ \ }\twoheadleftarrow\text{\ \ \ }\widehat{G\times
\mathcal{R}}\text{ \ \ \ \ }\hookleftarrow\text{ \ \ \ }\widehat{\mathcal{R}}%
\end{array}
,
\]
\vskip10pt \noindent where $\mathcal{O}_{d}:=C^{\ast}(\{\psi_{i},\psi
_{j}^{\ast}\})$ is a Cuntz algebra consisting of $d$-isometries $\psi_{i}$
($i=1,2,\cdots,d$): $\psi_{i}^{\ast}\psi_{j}=\delta_{ij}\mathbf{1}$,
$\overset{d}{\underset{i=1}{\sum}}\psi_{i}\psi_{i}^{\ast}=\mathbf{1}$.

The crucial ingredients for this scenario are as follows:

1) The essence of transitions from \textit{\textbf{invisible Micro}} with
\textit{\textbf{dynamical motions}} into \textit{\textbf{visible Macro}}
equipped with \textit{\textbf{universal indices}} can be found physically and
typically in the processes of \textit{\textbf{condensation}} (to form
condensed states), whose mathematical expression is nothing but the so-called
\textquotedblleft B-construction\textquotedblright\ (or \textquotedblleft
bar-construction\textquotedblright, \textquotedblleft basic
construction\textquotedblright, etc., with variety of names) and/or
\textquotedblleft heat kernel method\textquotedblright\ to extract topological
and/or homotopical invariants forming a \textit{\textbf{classifying space }%
}and playing the universal roles in \textit{\textbf{classifying objects}} in
question. Such a classical object as $G/H$ to classify degenerate vacua plays
universal roles in the sector structure of SSB in such a form as the base
space of the sector bundle, $\hat{H}\hookrightarrow\overset{\frown}%
{\widetilde{\mathcal{F}}^{H}}=G\underset{H}{\times}\hat{H}\twoheadrightarrow
G/H$.

2)\ The notions of \textit{\textbf{sectors}} or \textit{\textbf{pure phases}}
and \textit{\textbf{mixed phases}} have been introduced to clarify the mutual
relations between quantum Micro and classical Macro \cite{Unif03}. For this
purpose, we need classify representations of the algebra of physical variables
on the basis of \textit{\textbf{quasi-equivalence}} which is just the unitary
equivalence \textit{\textbf{up to multiplicities}}. The minimal units in this
classification are \textit{\textbf{factor states}} or \textit{\textbf{factor
representations}} whose \textit{\textbf{centres}} are trivial, and they are
called \textit{\textbf{sectors}} mathematically or \textit{\textbf{pure
phases}} physically. If a state or its GNS representation is not a factor
because of its non-trivial centre, it is called a \textit{\textbf{mixed
phase}} which can be canonically decomposed into sectors or pure phases.

3)\ In the context of measurement processes, the above Micro $\Longrightarrow$
Macro transitions are taking place in the \textit{\textbf{amplification
process}} to magnify the microscopic changes of quantum states at the contact
points between the systems and the microscopic ends (= probe systems)\ of
measuring apparatus into the macroscopic motions of measuring pointers
\cite{MicroMacro, Amp06}. In these papers, the process of this sort is shown
to be formulated as a L\'{e}vy process with (or, ideally without) small
deviations from it. What is most important in this process is the transitions
from a \textit{\textbf{mixed phase }}as a \textit{\textbf{virtual
probabilistic mixture}} of many different sectors or phases into their
\textit{\textbf{spatial configurations}} in the \textquotedblleft%
\textit{\textbf{real space}}\textquotedblright, each subdomain (= pointer
position)\ of which is occupied by a single sector or phase. In the context of
measurements, this phase separation is allowed to take place chronologically
as is indicated by the Born statistics rule, whereas it can occur
\textit{\textbf{spatially or synchronically}} in some such thermal contexts as
non-equilibrium states with certain degrees of stability.

4) The above problem of \textit{\textbf{phase separation }}from the physical
viewpoint can be viewed logically as the localized selections of the single
truth value (in the sense of standard two-valued logic) from a
\textit{\textbf{multi-valued logic }}(in the context of Boolean-valued
analysis of \textit{\textbf{probability}} space). This process can be
controlled by a logical method called \textquotedblleft%
\textit{\textbf{forcing}}\textquotedblright\ \cite{Cohen} (famous for
P.Cohen's use of it in the context of proving independence of continuum
hypothesis from the ZF axioms of set theory), resulting in a
\textit{\textbf{topos of sheaves}} on the \textit{\textbf{classifying space}}
$G/H=\mathcal{R}$ (consisting of degenerate vacua) whose core member is given
by a sheaf $\Gamma(G\underset{H}{\times}\widehat{H})$ of sections of the
sector bundle $\hat{H}\hookrightarrow\overset{\frown}{\widetilde{\mathcal{F}%
}^{H}}=G\underset{H}{\times}\hat{H}\twoheadrightarrow G/H$ on $G/H=\mathcal{R}%
$ describing the sector structure of the algebra $\widetilde{\mathcal{F}}^{H}$
of extended observables in terms of its factor spectrum $\overset{\frown
}{\widetilde{\mathcal{F}}^{H}}$. This is to be put in parallelism with the
\textit{\textbf{sheaf}} $\mathcal{R}\longmapsto E_{\mathcal{A}(\mathcal{R})}$
of local states in DHR sector theory, which means that the notion of local
states $E_{\mathcal{A}(\mathcal{R})}$ of a local algebra $\mathcal{A}%
(\mathcal{R})$ in a spacetime domain $\mathcal{R}$ should be understood to
correspond to a choice of a family of degenerate vacua in $G/H=\mathcal{R}$
arising from SSB, namely, to the context of considering states of extended
observables $\widetilde{\mathcal{F}}^{H}$ in reference to each member of
degenerate vacua belonging to $G/H$. This parallelism clearly shows the
existence of quantum fluctuations inside of each space(-time) point $x\in
G/H=\mathcal{R}$, to which extent space(-time) points are highly non-trivial
physical objects!!

5) The differences in the degrees of stability mentioned in 3) may be related
in a meaningful way to the corresponding differences in changeability between
some items to be put one place to another and the certain stable behaviours of
the \textquotedblleft names\textquotedblright\ attached to specific places. To
systematize such degrees of stability may be quite relevant for satisfactory
understanding of the various stabilized domains appearing in different levels
in nature and also of hierarchical structures of biological organisms, from
the viewpoint of Grothendieck's topoi and sites.

Last but not least!: I have benefited very much about the problems related to
the forcing from discussions with Mr. H. Saigo and Mr. R. Harada, to whom I am
very grateful. I would like to express my sincere thanks to Profs.~Belavkin,
Khrennikov, Smolyanov and Volovich for their valuable comments and kind
encouragements at QBIC2010.

\end{document}